# Title: Theoretical Analysis of a Two-Dimensional Metallic/Semiconducting Transition-Metal Dichalcogenide $NbS_2$//$WSe_2$ Hybrid Interface

Authors: Zahra Golsanamlou[1], Luca Sementa[1], Teresa Cusati[2], Giuseppe Iannaccone[2], Alessandro Fortunelli[1,*]

[1]CNR-ICCOM & IPCF, Consiglio Nazionale delle Ricerche, via G. Moruzzi 1, 56124, Pisa
[2]Dipartimento di Ingegneria dell'Informazione, Università di Pisa Via G. Caruso 16, 56122, Pisa, Italy.

**Abstract**

We report a first-principle theoretical study of a monolayer-thick lateral heterostructure (LH) joining two different transition metal dichalcogenides (TMDC): $NbS_2$ and $WSe_2$. The $NbS_2$//$WSe_2$ LH can be considered a prototypical example of a metal ($NbS_2$)/semiconductor($WSe_2$) two-dimensional (2D) hybrid heterojunction. We first generate and validate realistic atomistic models of the $NbS_2$//$WSe_2$ LH, derive their band structure and subject it to a fragment decomposition and electrostatic potential analysis to extract a simple but quantitative model of this interfacial system. Stoichiometric fluctuations models are also investigated and found not to alter the qualitative picture. We then conduct electron transport simulations analyze them via band alignment analysis. We conclude that the $NbS_2$//$WSe_2$ LH appears as a robust seamless in-plane 2D modular junction for potential use in opto-electronic devices going beyond the present miniaturization technology.

[*] Corresponding author, email: alessandro.fortunelli@cnr.it

# Introduction

Two-dimensional materials (2DMs) [1], made up of few-atom-thick layers of transition metal dichalcogenides (TMDC), are attracting increasing interest in both science and technology for their peculiar charge transport and optical response properties which promise to be exploitable e.g. in future optoelectronic devices [2]. The number of studies in 2DMs based on TMDC has increased exponentially since the first preparation of atomically thin exfoliated sheets of TMDC (such as $MoS_2$, $NbSe_2$ or $WS_2$), and especially after their successful synthesis via chemical vapor deposition (CVD) [3]. Although the detailed growth mechanism is still matter of debate [4-6]), soon afterwards CVD synthesis protocols were extended to composite materials, enabling the synthesis of both vertical (VH) and lateral (LH) heterostructures joining two different TMDCs, and the exploration of their transport and opto-electronics features [7-10]. Different combinations of LHs and VHs of 2D-TMDCs have therefore been prepared and subject to extensive research. LHs, in particular, offer the possibility of investigating and exploiting seamless and atomically sharp in-plane heterostructures, realizing the ultimate thickness limit for conducting/semiconducting junctions, in principle suitable to build an integrated circuit technology competitive, at the fundamental level of control of the electrostatic potential in the device, with present silicon-based technology [11, 12]. At the nanometer scale, first-principles quantum-mechanical (QM) simulations with atomistic detail are needed to predictively model fundamental interactions and the basic behavior of single material and interfaces. The first step of a proper multi-scale approach must therefore start from QM modeling and simulations [13-15]. Indeed, multi-scale simulations have shown to be able to predict transmission properties with a comparable accuracy with respect to experiment [16, 17]. Such a predictive ability can be combined with the precise knowledge of the atomistic structure of the investigated systems, and therefore the possibility of obtaining accurate information on both ideal (non-defective) and non-ideal (defective) systems [18], so as to give information on the maximum achievable value of any given property, as well as, indirectly, information on the level of defectivity of the experimental system [19]. More in general, accurate data on well-defined configurations as provided by computational studies lay the foundation of structure/property relationships, especially when analysis tools are developed that allow one to understand and interpret in depth both computational and experimental results [17].

Experimentally, Zhang et al. studied edge effects in $MoS_2$//$WSe_2$ monolayer/bilayer lateral heterojunctions [20], and determined the alignment of the valence band maximum (VBM) and conduction band minimum (CBM) of the two semiconductors at the LH. Li et al. were able to grow devices based on $WS_2$//$WSe_2$ LHs and explored sulfurization effects [21], finding that the replacement of W-Se by W-S bonds at the interface increased the system stability, improved the efficiency of electron transport, and decreased the Schottky barrier. The Schottky barrier was also measured for the $NbS_2$//$MoS_2$ VH heterostructure by Shin et al. [22], studying in particular how the doping concentration affects Schottky barrier height and the system's transport properties.

Same-material heterostructures (i.e., two different phases of the same stoichiometrically identical material) have been studied at both experimental and theoretical levels for 2H/1T (or 1T') lateral interfaces [23-25], revealing a consistency between theoretical and experimental pictures, such as the numerical values of the Schottky barriers or more exotic topological-insulating phenomena [26]. However, it has also been shown that the influence of the contact metal electrode should be considered in some cases to reconcile theory and experiment [23].

At the theoretical level, density-functional theory (DFT) has been employed to study several TMDC heterostructures, such as in-plane $WS_2//WSe_2//MoS_2$ LH [27]. Some studies used nanoribbon models of TMDC LHs (e.g., $MoS_2//WS_2$ LH nanoribbons with 14.6 Å width and different lengths [28]), and analyzed their electronic structure and transport properties, finding a strong dependence of transport on the configuration of the edges, whether zig-zag or arm-chair, in keeping with the fact that charge transport was found to mostly occur along the edges [28-30]. Ding et al. investigated metal/metal $NbS_2$//black-phosphorus and $NbS_2$//metallic(T-phase)-$WSe_2$ VHs [31], finding that the high work function of $NbS_2$ stabilizes p-type ohmic contact for both VHs. A comparative charge analysis suggested the importance of a sufficiently large work function in one electrode to compensate the reduction in work function due to interfacial dipoles. $MoS_2//WS_2$ and $MoSe_2//WSe_2$ VH systems were studied by Zhu et al. [32], who rationalized the evolution of the k-resolved band structure of these systems for different stacking configurations in terms of the interplay between orbital splitting and charge transfer effects. Cao et al. [33] focused on $MoS_2//WSe_2$ LH and on the influence of stoichiometric randomization at the interface. A limited number of stoichiometric swappings between S and Se atoms was found to have a minor effect on the electrostatic potential profile at the junction, whereas a more extensive defectivity level was found to smoothen the electrostatic potential profile, reducing the ability of the device to promote exciton dissociation. Jelver et al. calculated Schottky barrier height of 2D metal-semiconductor junctions [16], considering the slope of the density of states (DOS) and estimating barrier height from values of transmission. They also studied how a higher doping level and the presence of localized interface states can modulate the barrier height.

Here we conduct a first-principle theoretical study of a $NbS_2//WSe_2$ LH as a prototypical example of metal/semiconductor hybrid heterostructures. After generating realistic atomistic models of the $NbS_2//WSe_2$ LH and of its band structure, we use fragment decomposition and electrostatic potential analysis to arrive at a physically simple and quantitatively precise modeling of this interphase. Explicit transmission simulations confirm the expectations of our modeling, with consistency further demonstrated via a band alignment analysis of the transmission results. We conclude that the $NbS_2//WSe_2$ LH appears as a robust and potentially useful modular junction for use in opto-electronic devices. This is further supported by investigating stoichiometric fluctuations at the interface, corresponding to stoichiometrically non-sharp interfaces possibly produced at the experimental level, which shows that such phenomena can indeed occur but should not alter qualitatively (or even reinforce) the robustness and potential use of this system.

The article is organized as follows. The computational method and analysis protocol are presented in Section 2. Section 3 is devoted to presenting and discussing our main results: structure prediction, band alignment, transmission simulations and their analysis, stoichiometry fluctuations. Conclusions are summarized in Section 4.

## 2. Computational and Theoretical Method

### 2.1 Computational Details

Geometry optimizations and electronic structure calculations were carried out within first-principle density functional theory (DFT) using the Quantum Espresso (QE) package [34, 35]. A plane-wave basis set, a gradient-corrected exchange-correlation functional [Perdew-Burke-Ernzerhof (PBE)] [36] augmented with terms describing dispersion interactions in the Grimme-D3 formalism [37], and scalar-relativistic ultrasoft pseudopotentials (US-PPs) were utilized. Pure $NbS_2$ and $WSe_2$ hexagonal phases were described for convenience using orthogonal unit cells. Monkhorst-Pack k-meshes of $12\times 22\times 1$ and $1\times 22\times 1$ were used for the unit cell and the scattering region, respectively, to sample the Brillouin zone (BZ). We used an energy cutoff of 50 Ry for the selection of the plane-wave basis set to describe the wave function and 500 Ry for the electron density, and a vacuum of 22 Å was used to minimize interactions with replicated unit cells.

Transmission simulations were performed using the PWCOND routine [38, 39] within QE, based on a scattering state approach which integrates numerically a scattering equation in real space along the direction of transport. In this case the initial DFT (QE/PWscf) calculations were carried out using 80 Ry as energy cut off and 800 Ry as the electron density cutoff. In PWCOND we also used 40 k-points in the y direction (direction perpendicular to transmission) of the orthogonal unit cell, checking that results were converged up to 110 k-points.

### 2.2 The Analysis Method: Fragment Decomposition and Electrostatic Potential Descriptor

To achieve a deeper understanding of DFT and transport simulations we employ a basic approach for analyzing the electronic band structure of complex systems as proposed in Ref. [17], here slightly refined as described hereafter and schematically illustrated in Figure 1. This approach relies on two main pillars: (1) decomposing a complex system (here: the LH) in terms of individual fragments (here: the $NbS_2$ and $WSe_2$ extended phases), and thus exploiting the information available on such simpler systems to quantitatively dissect and predictively understand the behavior of the complex system, and (2) using the electrostatic potential as a basic and unifying descriptor of transport phenomena.

In detail, we first conduct DFT band structure simulations on a given LH atomistic model, named the "scattering system", see Figure 1a. We then select two fragments in this scattering system: (i) one corresponding to a $NbS_2$ (orthogonal) unit cell, and (ii) one corresponding to a $WSe_2$ (orthogonal) unit cell, and we replicate them to produce $NbS_2$ and $WSe_2$ extended phases, see Figure 1(d) and Figure 1(h), respectively. Note in passing that the lattice parameter in the y direction resulting from relaxation is a tradeoff between the lattice parameters of $NbS_2$ and $WSe_2$ extended phases, and it is important to consider such average geometry in the leads to get an accurate electrostatic potential analysis, see the discussion below. We recall that $NbS_2$ is metallic as apparent from its band structure (Figure 1(c)), while $WSe_2$ is a semiconductor (see the band structure in

Figure 1(g). On such pure-phase fragments we calculate the electrostatic potential from Nb and W atoms to the vacuum, see the plots in Figure 1(b) and Figure 1(f), and draw basic fragment quantities, i.e.: (i) for $NbS_2$ the difference between the on-site electrostatic potential on the Nb atom ($V_{Nb}$) and the Fermi energy ($E_f$) which we name $\Delta_{E_f/Nb}$, Figure 1e, (ii) for $WSe_2$ the difference between the on-site electrostatic potential on the W atom ($V_W$) and the top of the valence band (TVB) which we name $\Delta_{TVB/V_W}$. For the scattering system, we then calculate the on-site electrostatic potential on the Nb and W atoms, Figure 1(m). Finally, we merge fragment and complex system analysis by adding the $\Delta_{E_f/Nb}$ and $\Delta_{TVB/V_W}$ fragment quantities to the on-site electrostatic potentials of the scattering system to obtain the band alignment, i.e., the local Fermi energy for the metal $NbS_2$ component, Figure 1(j), and the local top of the valence band for the semiconductor $WSe_2$ component, Figure 1(k). The final result is illustrated in the plot of Figure 1(l).

In passing, we note that the electrostatic potential is the main component of the Kohn-Sham potential, which provides the self-consistent background on which electrons move. What is lacking in the electrostatic potential with respect to the Kohn-Sham one are exchange-correlation terms. However, since such terms are evaluated via a numerical quadrature in DFT codes, we have found it numerically more robust to employ the bare electrostatic potential in our analysis [40]. It can be recalled that all such quantities as the electrostatic potential, the Kohn-Sham potential, etc., rigorously correlate via the Hohenberg-Kohn theorem, so that any of them can be used indifferently. Also, the choice of the site where to calculate the electrostatic potentials is in principle irrelevant: we chose on-site atomic electrostatic potentials (i.e., on the sites of the nuclei) because they are easier to identify precisely from the atomic coordinates. Finally, it can be noted that the fragments must be taken far enough from the interfaces to avoid interface effects on the analysis: to this end we performed convergence tests by increasing the size of the scattering region from a 4+4 to an 8+8 system (where N+N indicates the number of orthogonal unit cells included) until we got a flat behavior of the on-site potentials in the middle of the $NbS_2$ metal component.

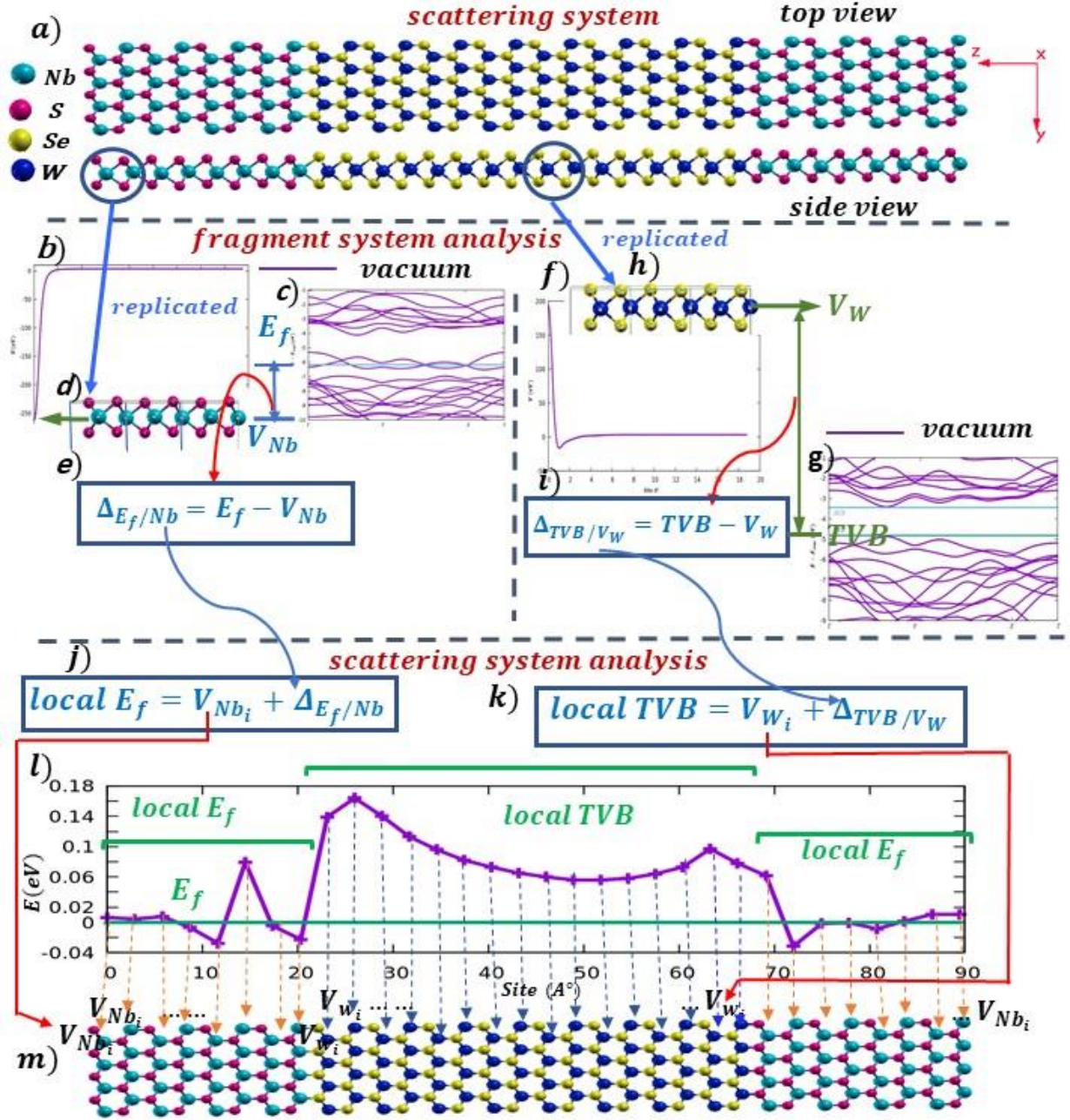

***Figure 1.*** *Analysis of the electrostatic potential for the NbS$_2$//WSe$_2$ LH: a) top and side view of LH with definition of fragments, (b-d) NbS$_2$ fragment: b) electrostatic potential from the Nb nucleus into the vacuum, c) band structure, d) replicated NbS$_2$ fragment (side view), e) $\Delta_{E_f/Nb}$ = difference between the on-site potential on the Nb atom and the Fermi energy ($E_f$) of the NbS$_2$ fragment; (f-i) WSe$_2$ fragment: f) electrostatic potential from the W nucleus into the vacuum, g) band structure, h) replicated WSe$_2$ fragment (side view), i) $\Delta_{TVB/V_W}$ = difference between the on-site potential on W atom and the top of the valence band (TVB) of the WSe$_2$ fragment, j) the local Fermi level for the NbS$_2$ component of the scattering system is obtained by combining the Nb onsite potential in scattering system with the $\Delta_{E_f/Nb}$ NbS$_2$ fragment quantity obtained in e), and j) the local top of the valence band (TVB) for the WSe$_2$ component of the scattering system is obtained by combining the W onsite potential with the e $\Delta_{TVB/V_W}$ WSe$_2$ fragment quantity obtained in i); (m-l) the atomistic model m) is depicted again and the final band alignment l) is plotted.*

## 3. Results and discussion

### 3.1 Structure generation

To generate an atomistic model of a $WSe_2//NbS_2$ LH, we first validated the accuracy of our DFT approach in terms of prediction of the geometrical features of $WSe_2$ and $NbS_2$ monolayer (ML) phases. The lattice parameters from previous-theory for $WSe_2$ and $NbS_2$ monolayer (ML) phases are: 3.319 $A°$ and 3.344 $A°$, respectively, [database-website], very close to our DFT-predicted (full DFT geometry relaxation including cell axes) are: 3.323 $A°$ and 3.346 $A°$, respectively. We then built up 3 systems of increasing size by matching (N+N) replicated orthogonal unit cells of $WSe_2$ and $NbS_2$, where N is the number of unit cells of each component, and subjected the resulting system to full DFT geometry relaxation (relaxation included cell axes) and band structure and electrostatic potential analysis. We considered 4+4, 6+6 and 8+8 $WSe_2//NbS_2$ systems, and the 8+8 system was selected for transmission analysis because the electrostatic potential profile was tested to be at convergence, as discussed in the Supplementary Information, Figure SI.2. The atomistic model outcome of this step is depicted in Figure SI.1. It should be noted that the mismatch between the hexagonal symmetry of the ML phases and the orthogonal unit cells needed for the successive transport simulations makes that we have 2 slightly different $WSe_2//NbS_2$ interfaces on the right- and left-hand-side.

### 3.2 Electrostatic potential analysis

We apply the analysis scheme of Section 2.2 to the 8+8 $WSe_2//NbS_2$ LH system (see Figure1(a)). The resulting band alignment, i.e., the local Fermi energy ($E_f$) for the metal $NbS_2$ component and the local top of the valence band (TVB) for the semiconductor $WSe_2$ component, is reported in Figure 2. In keeping with the fact that $NbS_2$ is a metal, we find a nearly flat local Fermi energy far from the interfaces in Figure 2, which is consistent with a zero electric field and potential gradient. Also the local TVB on the W atoms is flat far enough from the interfaces. It can be noted that a flat local $E_f$ in $NbS_2$ far from the interfaces is needed to choose these sites as leads of the transmission simulations of Section 3.3 and to allow for a precise projection of the wave function of the scattering system onto the wave function of the reference system. It can also be noted the asymmetric profile of the local TVB indicating that there are two slightly different interfaces in the chosen LH, due to geometric non-equivalence reasons. The two interface dipole moments must however be equal in absolute value and opposite in sign so as to cancel each other, since we are working with plane-wave codes and periodic models (see the discussion in the SI section 2, for the 4+4 $WSe_2//NbS_2$ LH system, Figure SI.3 and Figure SI.4). These effects are anyway minor, and we did not further investigate them.  It is instead worthwhile noting that the local Fermi energy in the middle of the unit cell coincides with the system Fermi energy up to a hundredth of an eV proving the accuracy of the chosen analysis strategy. Also, it can be noted that there is a difference between the system work function in the fully relaxed and fragment unit cells of pure $NbS_2$ amounting to 0.15 eV, while the difference between the local TVB in relaxed and fragment unit cells of pure $WSe_2$ is 0.11 eV (see the SI), implying some long-range relaxation effects in a mesoscopic LH (we leave investigation of these effects to a future multi-scale investigation). To provide more details, Figure SI.5 compares the DOS of fragments taken from the scattering region with the fully relaxed unit cells of $WSe_2$ and $NbS_2$ monolayer systems.

Finally, as a main outcome of our analysis, we single out the jumps in the local $E_f$ and TVB at the interfaces due to interfacial potential gradients. We quantify these jumps to be of the order of 0.1-0.2 eV, providing an estimate of the Schottky barrier (see below). Note that the Schottky barrier is negative in this system.

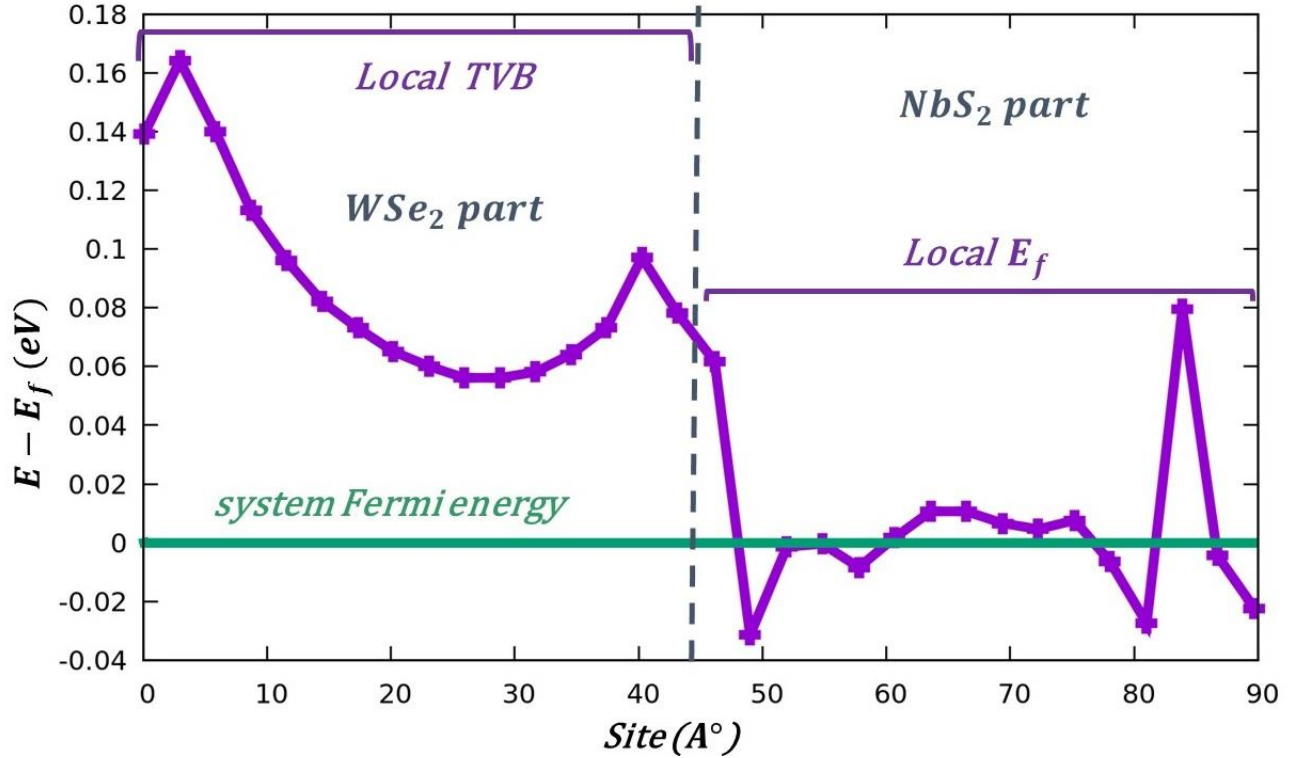

**Figure 2.** *Local top of the valence band (TVB) on the W atoms and Local Fermi energy ($E_f$) on the Nb atoms of the 8+8 $WSe_2//NbS_2$ LH.*

### 3.3 Transmission Simulations

We perform transmission simulations for the scattering system, choosing a $NbS_2$ orthogonal unit cell far from the interfaces as a left lead, followed by 3 $NbS_2$ unit cells, 8 $WSe_2$ unit cells, 3 other $NbS_2$ unit cells and and one $NbS_2$ orthogonal unit cell as the right lead, as shown in Figure 1(a). As reference systems, we perform band structure calculations of $NbS_2$ and $WSe_2$ fragments using the geometry extracted from the relaxed geometry of the scattering system. We recall that the cell dimension in the y direction (perpendicular to the transport direction) does not coincide with the equilibrium one of either $NbS_2$ or $WSe_2$ pure ML extended phases, but it is a compromise between the two as obtained from the relaxation of the $WSe_2//NbS_2$ LH, and therefore analogously also the corresponding wave functions. The transport coefficient as evaluated by the PWCOND code from the left lead through the central region of the scattering system to the right lead is shown in Figure 3, and has a reasonably smooth profile with a sizeable maximum reading 0.25. The $WSe_2//NbS_2$ LH therefore appears to be a robust and suitable system for use in electronic devices.

Interestingly, we can use the density of states (DOS) of the pure phases and the band alignment profile of Figure 2 to deepen our analysis and find the energy intervals (the range of bias) in which we expect finite electron transport and even to obtain a quick estimate of the transmission coefficient in Figure 3, thus checking its consistency and physical interpretation.

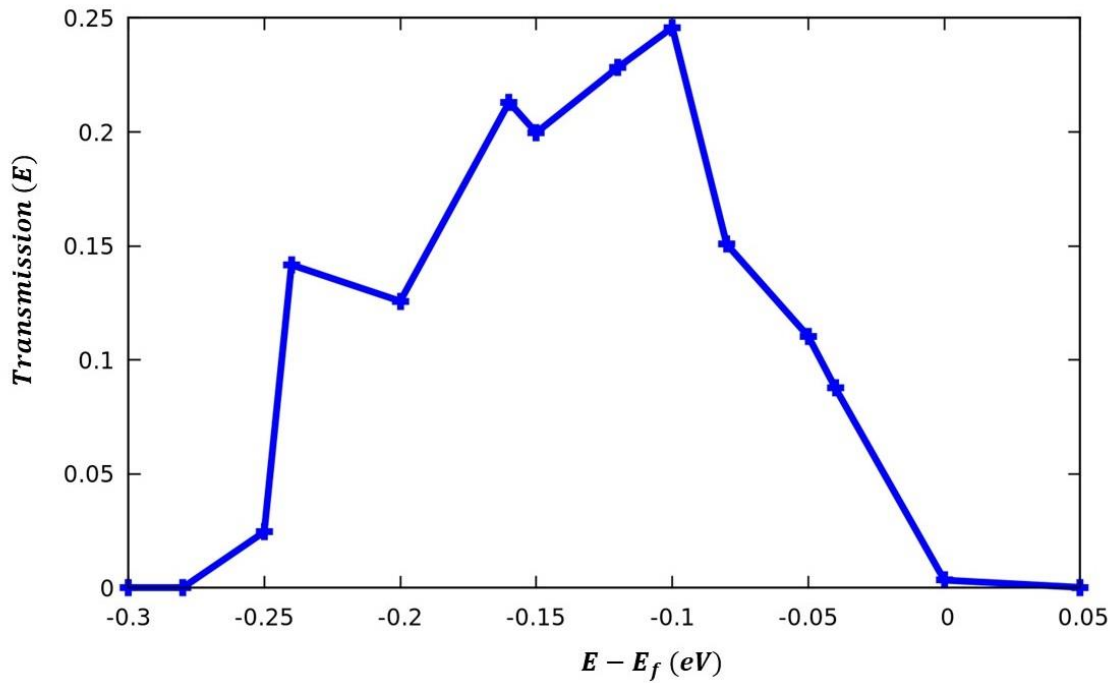

**Figure 3.** *The transmission coefficient as calculated for our $WSe_2//NbS_2$ LH model right below the Fermi energy.*

By convoluting in fact the band structure of the pure phases and the electrostatic potential profile, we can roughly predict the number of states available for transmission in each $NbS_2$ or $WSe_2$ units, thus finding e.g. where a zero transmission coefficient is to be expected. Figure 4 illustrates this analysis: we divide the monolayer phase DOS in a green and a yellow part to distinguish the region of higher density of states (in green) from the region of lower density of states (in yellow), and we align them along the transmission direction to identify values of bias in which we expect that there will not be enough states available for transmission from values of bias where we expect that there will be enough states for transmission. It is worthwhile noting that the limited energy width of the high DOS region in $NbS_2$ and especially the Nb atoms at the interface mostly affected by the interface dipole and whose local $E_f$ is therefore shifted to higher energies are the ones responsible for cutting the low-energy end of transmission. Figure 4 pictorially illustrates how our analysis in terms of electrostatic potential, local $E_f$, etc. turns out to be useful and effective, allowing us to rather precisely single out the range of finite transmission coefficient. We repeat our main conclusion that transmission turns out to be sizeable in an interval of ≈0.28 eV below the Fermi level, reaching a maximum of 0.25.

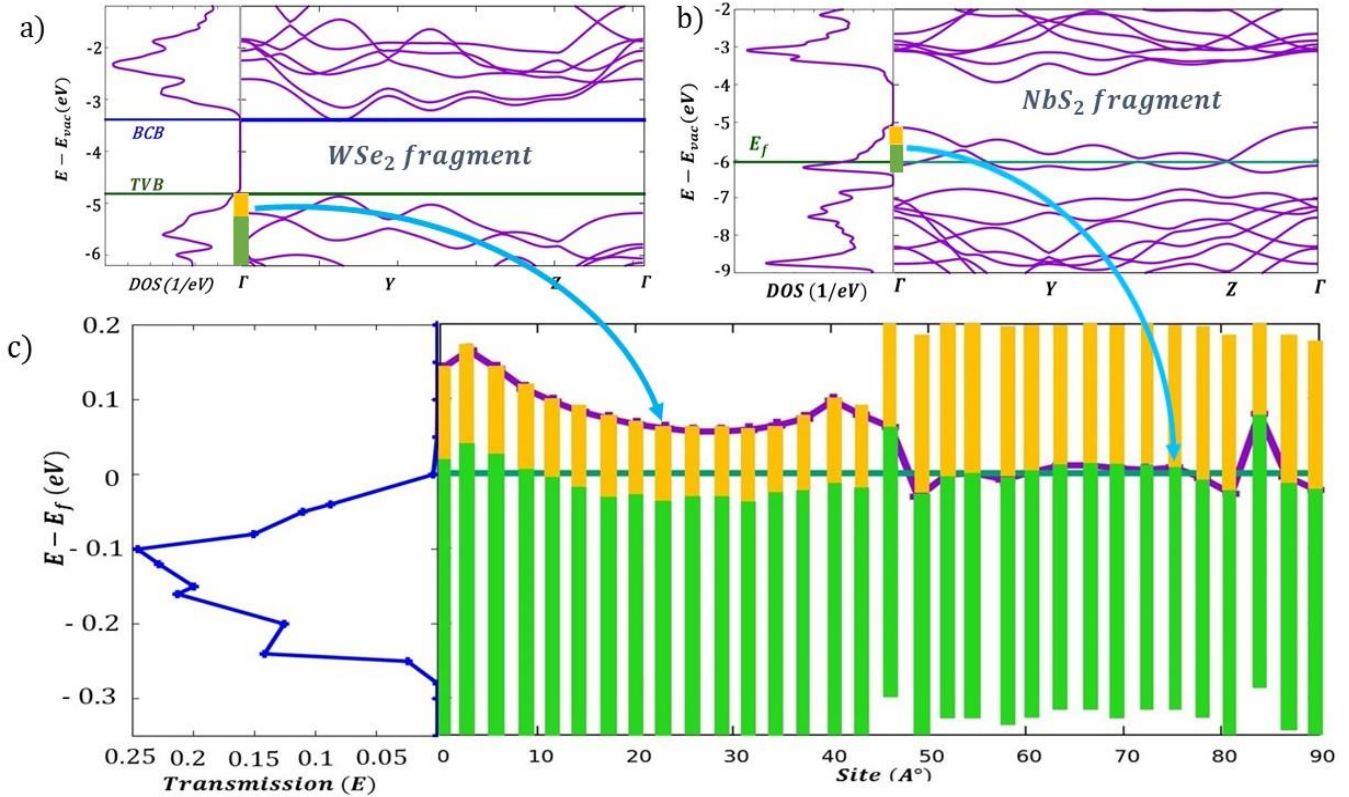

**Figure 4.** *Our approach to find the nonzero energy interval for transmission analysis: a) DOS and band structure of $WSe_2$ fragment, b) DOS and band structure of $NbS_2$ fragment, c) transmission of $WSe_2//NbS_2$ LH and band alignment procedure to find the states for electron transmission using the obtaind color bar from a) and b) in place of local $E_f$ and TVB.*

### 3.4 Stoichiometry fluctuations at the LH interface

Since in the investigated $WSe_2//NbS_2$ LH two different chalcogenide atoms are present on the two sides of the interface, we considered the possibility that imperfect, stoichiometrically non-sharp interfaces are in fact produced at the experimental level. We explore this possibility differently from the constant stoichiometry simulations of GRef. [33]. Instead, we exchange rows of S atoms with Se ones or vice versa, and predict the energetics associated with this process and its effect on the on-site electrostatic potentials. From the energy point of view, in general we find that the structure in which S atoms replace Se atoms are the most stable ones. We study the electrostatic potential on the Nb and W atoms to see how much this exchange can affect the physical properties of the LH.

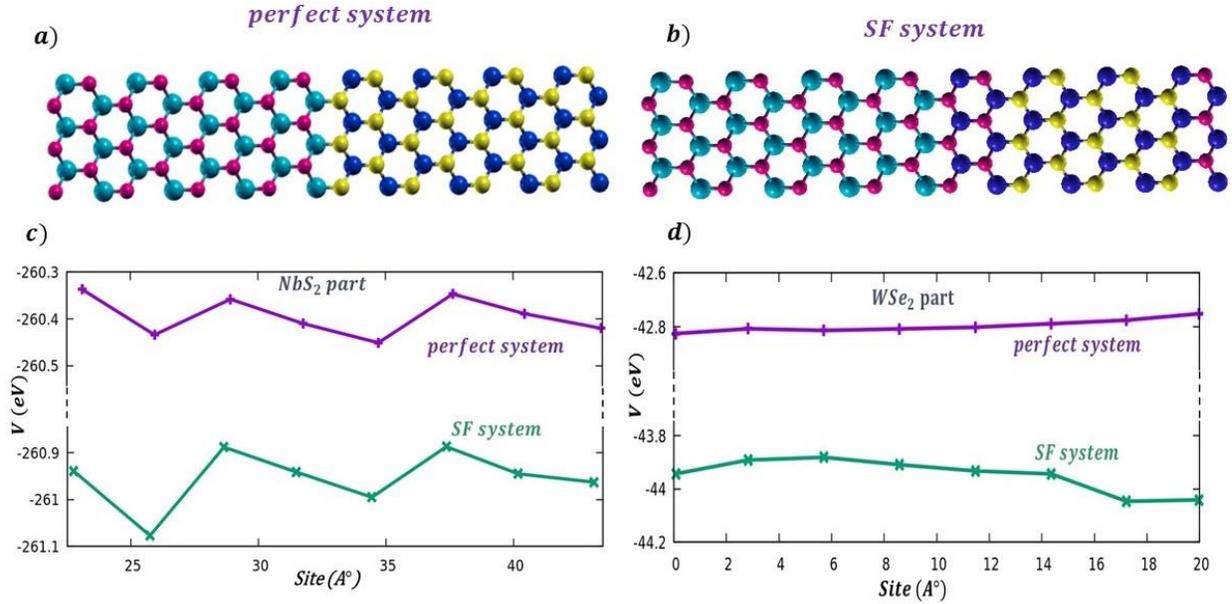

**Figure 5.** *Stoichiometry fluctuation at the interfaces of 4+4 $WSe_2//NbS_2$ LH : a) perfect system structure, b) stoichiometry fluctuated (SF) system, c) comparison between potential on Nb atoms in the perfect and SF systems, d) comparison between potential on W atoms in the perfect and SF systems.*

The difference between on-site potential of Nb atoms in perfect 4+4 $WSe_2/$/$NbS_2$ LH is in the range 0.031-0.104 eV and for $W$ atoms is between 0.0057-0.024 eV. Also, the difference for Nb atoms of stoichiometry fluctuated (SF) system is 0.018-0.138 eV, while for W atoms it is 0.0054-0.0516 eV. The jump in the electrostatic potential is therefore small: it can change a little the Schottky barrier and the band alignment however without leading to a drastic expected change in transmission. Moreover, the increased thermodynamic stability of the system also likely protects it against formation of other defects, therefore leading to improved transmission properties as observed in experiment [21].

## 4. Conclusions

Two-dimensional [1] lateral heterostructures (LH) joining two different transition metal dichalcogenides (TMDC) materials stand out as promising modular unit blocks in future optoelectronic devices [11] due to their peculiar charge transport and optical response properties [2,3]. Several examples of such composite interfaces are currently being produced [7-10] and investigated at both experimental [20-22] and theoretical [27-33] levels, with the goal of realizing the ultimate thickness limit for metal/semiconductor in-plane junctions. At this nanometer scale, first-principles quantum-mechanical (QM) simulations are required to predictively model these systems and their behavior with an accuracy comparable with experiment [18], and are especially useful when analyzed to interpret in depth computational and experimental [19] results, and extract basic information on relationships between structure and transport properties [16, 17].

Here, we use QM modeling to investigate the electronic structure and transport features of a monolayer-thick $NbS_2$//$WSe_2$ lateral heterostructure (LH), as a paradigmatic example of metal/semiconductor 2D hybrid interfaces based on TMDC. We construct and validate convergence of atomistic periodic models of the LH, derive their band structure also exploring sulfur/selenium stoichiometric fluctuations at the interface, and predict electron transport response. An accompanying fragment and electrostatic-potential (plus band alignment) analysis illuminates our findings, and allows us to interpret them in terms of a simple but quantitative model of the junction. We find that the $NbS_2$//$WSe_2$ LH is a promising robust in-plane 2D modular junction, with stoichiometric fluctuations possibly stabilizing the system without deteriorating its electronic features, and thus appears suitable to be potentially exploited in an integrated circuit technology going beyond present silicon-based miniaturization level [12].

## Acknowledgements

We gratefully acknowledge financial support from the QUEFORMAL FET-Open EU project (Grant Agreement 829035) and from the MIUR PRIN Five2D project (2017SRYEJH), and computational support from the Cineca Supercomputing Center for the LH-FET grant. We are grateful to Alexander Smogunov for useful discussions and hints about transmission simulations and to Enrique González Marín for providing useful scripting.

# Supplementary Information for "Theoretical Analysis of a Two-Dimensional Metallic/Semiconducting Transition-Metal Dichalcogenide NbS$_2$//WSe$_2$ Hybrid Interface"

Zahra Golsanamlou[1], Luca Sementa[1], Teresa Cusati[2], Giuseppe Iannaccone[2], Alessandro Fortunelli[1,*]

[1]CNR-ICCOM & IPCF, Consiglio Nazionale delle Ricerche, via G. Moruzzi 1, 56124, Pisa
[2]Dipartimento di Ingegneria dell'Informazione, Università di Pisa Via G. Caruso 16, 56122, Pisa, Italy

Along research on two-dimensional (2D) materials [1, 2], and the experimental investigations on transition metal dicalcogeneides (TMD) and their heterostructures [3-12], previous theoretical studies have considered Lateral (LH) and Vertical Heterostructure systems based on transition metal dochalcogeneides typically using density functional theory (DFT) [13-15], focusing on magnetic/nonmagnetic ground states of LHs [16, 17], transmission and current-voltage characteristics [16], and structural defects [18, 19]. As described in the main text, our work follows these lines, and in this Supplementary Information we provide computational details and further information to complement that provided in the main text. In particular: (a) we report convergence tests to support the choice of the size of the scattering system, (b) we provide an estimate of the interfacial dipole via an analysis of the electrostatic potential in the vacuum, (c) we compare density of states (DOS) of WSe$_2$ and NbS$_2$ either fully relaxed or at the geometry taken from the scattering system for the transmission calculations, and (d) we report transmission coefficient away from the Fermi level to complement the plots reported in Figure 4 of the main text.

## 1) Tests of convergence of the electrostatic potential with the size of the system

To select a scattering system with a size appropriate to investigate accurately interfacial phenomena, we conducted tests on 4+4, 6+6, up to 8+8 WSe$_2$//NbS$_2$ LH systems, where N+N represents replicated orthogonal unit cells of WSe$_2$ and NbS$_2$, and N is the number of unit cells of each component. For these systems, we performed full DFT geometry relaxation (relaxation included cell axes, the initial lattice parameter was taken as an average of pure systems for the y direction), see example depictions in Figure SI.1.

---

[*] Corresponding author, email: alessandro.fortunelli@cnr.it

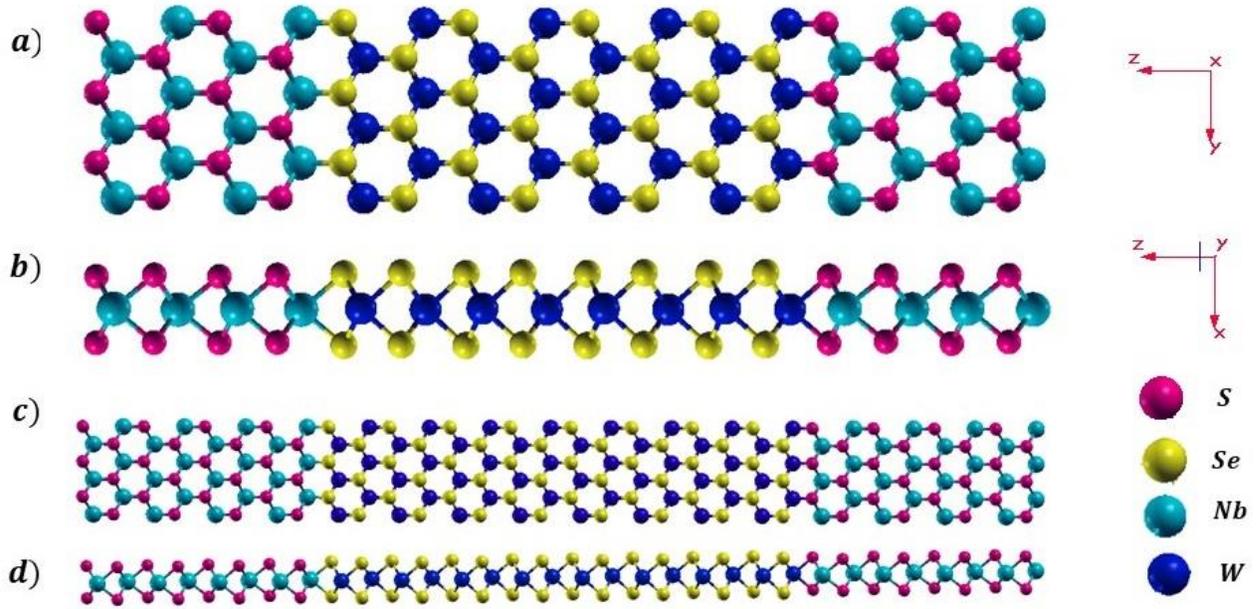

*Figure SI.1. Structure of $WSe_2//NbS_2$ LH: a) 4+4 LH, top view, b) 4+4 LH, side view, c) 8+8 LH, top view, and d) 8+8 LH, side view*

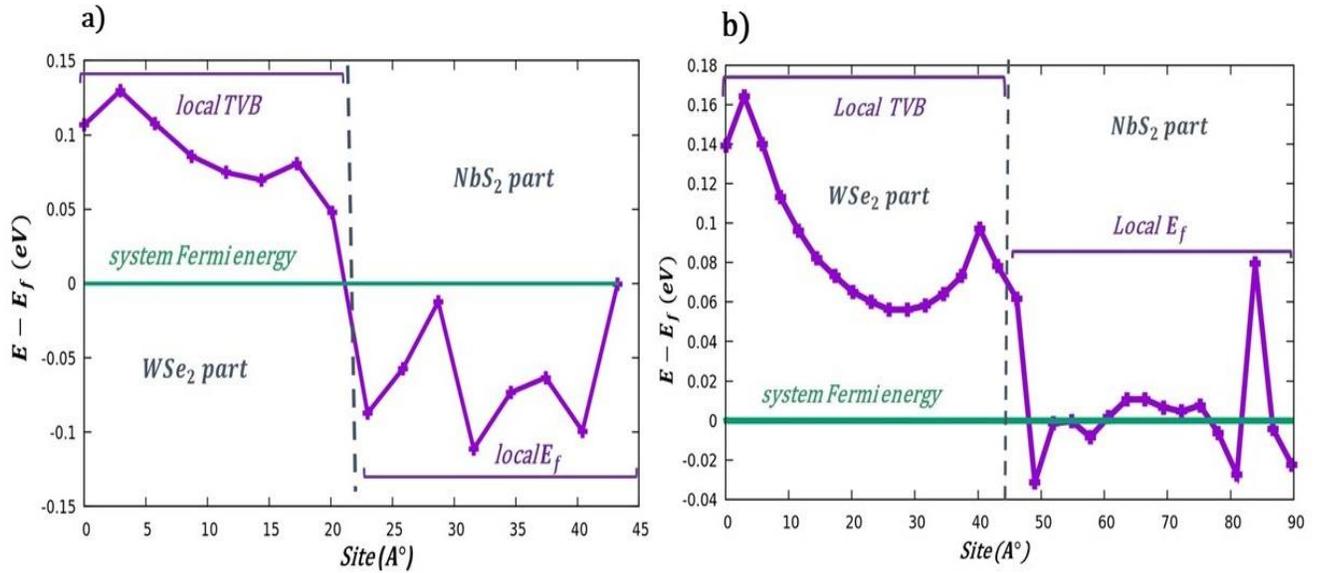

*Figure SI.2. Local top of the valence band (TVB) on the W atoms and Local Fermi energy ($E_f$) on the Nb atoms for: a) 4+4 $WSe_2//NbS_2$ LH, and b) 8+8 $WSe_2//NbS_2$ LH.*

Focusing for simplicity on 4+4 and 8+8 systems, our system are constituted by one $NbS_2$ orthogonal unit cell (then used as the left lead in the transmission simulations) followed by 1 (or 3) other $NbS_2$ unit cells for the 4+4 (or 8+8) case, 4 (or 8) $WSe_2$ unit cells in the middle of the scattering region, and one $NbS_2$ orthogonal unit cell (used as the right lead in the transmission simulations) positioned after 1 (or 3) other $NbS_2$ unit cells. Analogous construction holds for the 6+6 system.

We relaxed the geometry of these systems and calculated the on-site electrostatic potentials (i.e., the electrostatic potentials on the positions of Nb and W atoms), as shown in Figure SI.2a for the 4+4 case. In Figure SI.2a one can appreciate, apart from the expected jump of the local Fermi energy on the Nb atoms at the interface due to charge reorganization, distinct oscillations showing that the Nb atoms are not sufficiently far from the interface and are therefore still affected by the electric field created by the potential gradient. Also the local top of the valence band (TVB) on the W atoms in Figure SI.2a, naturally exhibiting a different behavior with respect to $NbS_2$ consistent with the fact that $WSe_2$ is a semiconductor, is not well converged far from the interfaces. We do not find for the 4+4 $WSe_2//NbS_2$ LH system the plateau in the local Fermi energy far from interfaces that is needed to perform transmission simulations and we discard this system. Analogously, we found that the 6+6 system (not shown) presents attenuated but still apparent convergence problems. On the contrary, for the larger 8+8 $WSe_2//NbS_2$ LH (shown in Figure SI.2b), the local Fermi energy on the Nb atoms far from the interfaces is approximately flat and the local TVB on the W atoms in the scattering region also reaches convergence in the middle of the unit cell. We then consider the 8+8 $WSe_2//NbS_2$ LH system large enough for further investigation.

2) **Estimate of interfacial dipoles**

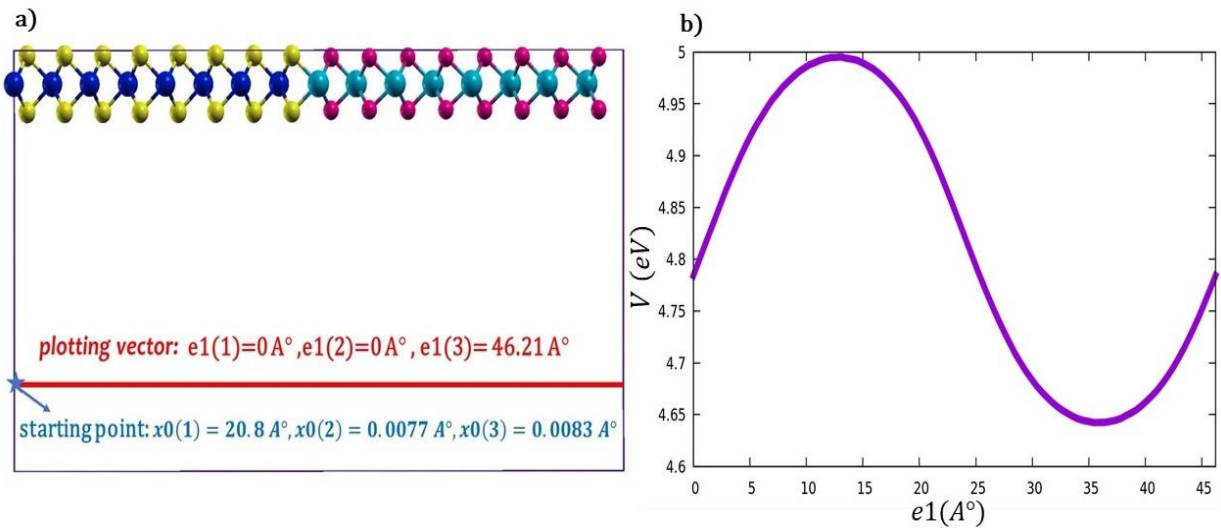

*Figure SI.3. Analysis of the electrostatic potential to assess interfacial dipoles: a) the side view of 4+4 $WSe_2//NbS_2$ LH, and b) the potential profile in the vacuum along the transmission direction (z axis)*

As mentioned in the main text, due to geometric reasons we have two different interfaces in $WSe_2//NbS_2$ LH systems, with potentially two different interfacial dipoles. However, in a periodic 3D approach as the one employed in this work the system cannot have a dipole (it would cause an electrostatic catastrophe) so that the dipoles at the two interfaces slightly adjust as to exactly cancel each other. To provide a quantitative estimate of these dipoles (also giving an estimate of the charge reorganization at the interface), we analyze the electrostatic potential in the vacuum outside the scattering system along z (the transmission direction). Assuming that the dipoles are localized at the interfaces, from a plot of the electrostatic potential in the vacuum we can derive the value of the dipoles by fitting the plot via a simple analytic expression for the potential produced by two dipoles equal in absolute value and opposite in sign located at the interfacial sites. The path for calculating potential is given by the red line in Figure SI.3a. The fit gives a dipole of 2.2 Debye at each interface (equivalent e.g. to 1.1 elementary opposite charges at a distance of 2.0 Å).

Another useful quantity which can be extracted from this type of plots is the value of the electrostatic potential in the vacuum, related the system work function. Note in fact that the potential in Figure SI.3b oscillates by ≈0.2 eV, which can be therefore considered as the order of magnitude of the uncertainty for the electrostatic potential in vacuum. To obtain a more accurate estimate we investigated "mirrored" systems, i.e., we built up models in which the $WSe_2//NbS_2$ LH system was mirrored around one point as shown in Figure SI.4a for the 4+4 $WSe_2//NbS_2$ LH. Mirroring the system exactly cancels dipolar terms (therefore we remove dipole corrections in the DFT calculation) so that only quadrupole contributions remain, as shown by the electrostatic potential 2D plot in Figure SI.4b. We can then take the potential in the exact middle of the unit cell as a more accurate value of the potential in the vacuum.

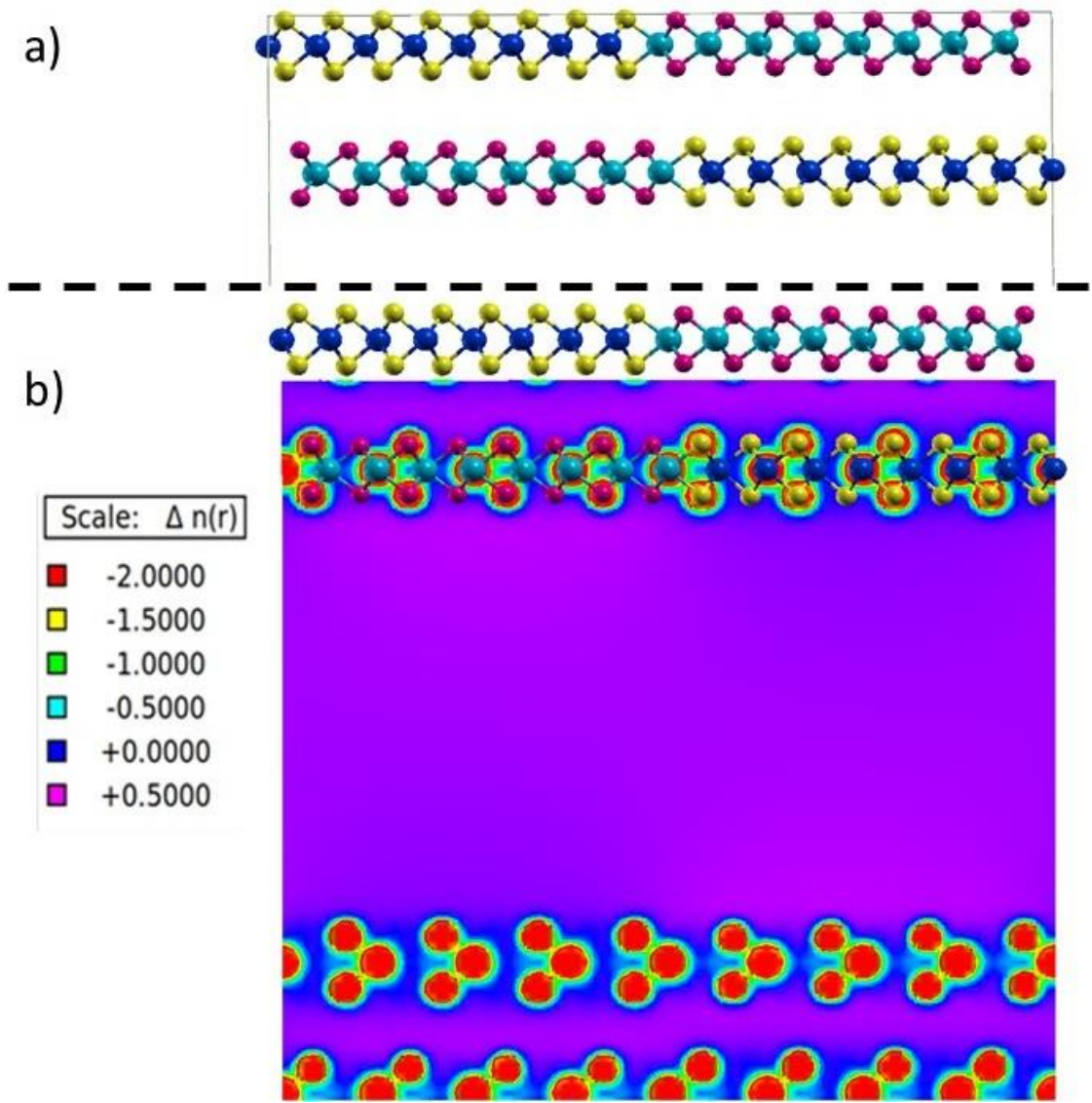

*Figure SI.4. 2D potential analysis of mirroring system: a) the side view of mirrored 4+4 $WSe_2//NbS_2$ LH, and b) the 2D potential profile which colors show the negative and positive potentials.*

### 3) Comparison between the DOS of $WSe_2$ and $NbS_2$ fully relaxed and geometry taken from the scattering region unit cells

As mentioned in the main text (Section 2.2), the geometry of the $NbS_2$ and $WSe_2$ fragments in the scattering system differ from the fully relaxed ones of the pure phases due to the constraints associated with the formation of the LH. As leads, we select $NbS_2$ and $WSe_2$ fragments (orthogonal unit cells) far from interfaces. We further process them by taking the *y*-parameter of these unit cells as in the scattering system and by relaxing the other geometrical parameters. Figure SI.5 then compares the DOS of $WSe_2$ and $NbS_2$ unit cells in the fully relaxed and scattering-system geometries. The TVB with respect to vacuum is -5.0378 eV for the fully relaxed WSe$_2$ unit cell, while it is -4.9374 eV for the fragment WSe$_2$ taken from the LH scattering system. The work function is 6.0757 eV for the fully relaxed NbS$_2$ unit cell, while it is 5.9184 eV for the NbS$_2$ fragment taken from LH. The upward shift is then 0.15 and 0.11 eV for NbS$_2$ and WSe$_2$, respectively. From such simulations we could also obtain the strain energy of each phase due to the formation of the interface, i.e., the energy difference between the fragment taken from the LH scattering system and the fully relaxed unit cell, but for these size-matched systems these effects are below 0.01 eV and we did not investigate them further.

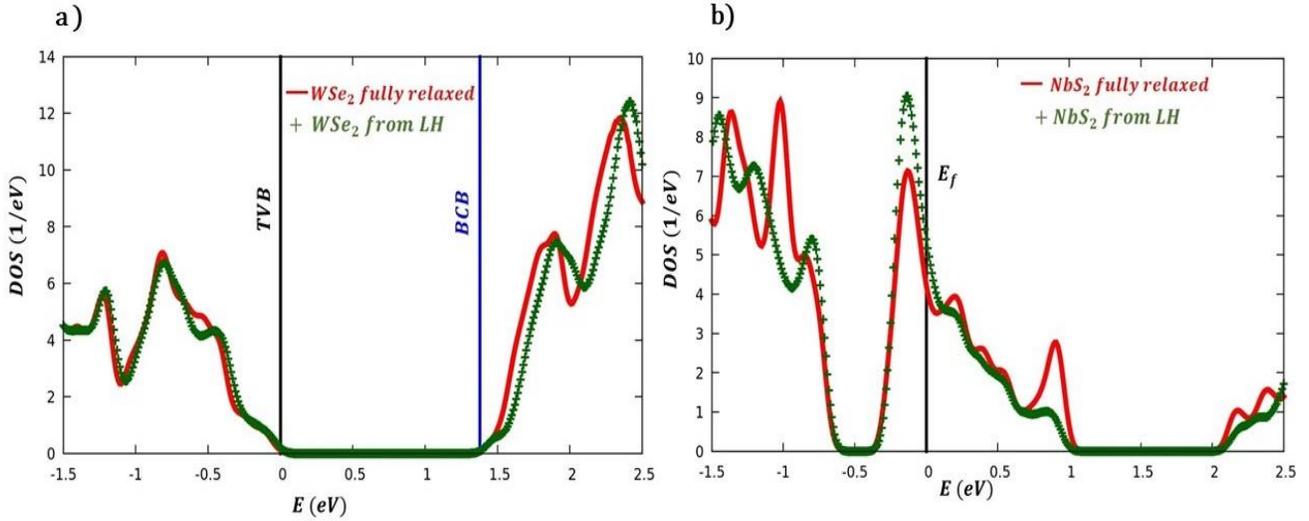

*Figure SI.5. DOS of fully relaxed and geometry taken from the scattering region of a) WSe$_2$ unit cells, where the fully relaxed TVB is -5.0378 and for fragment from LH is -4.9374 eV with respect to vacuum, and b) NbS$_2$ unit cells with $E_f = -6.0757$ eV fully relaxed and $E_f = -5.9184$ eV for fragment from LH with respect to vacuum*

### 4) Transmission simulation at a bias far from the Fermi level

For completeness and to complement the plot reported in Figure 4 of the main text, we report in Figure SI.5 the transmission coefficient for 8+8 $WSe_2//NbS_2$ LH calculated far from the Fermi level. The plots in Figure SI.5 are consistent with band alignment arguments discussed in the main text (Section 3.3) and the presence of a large number of states available for electron transport, as apparent by the high value of the transmission coefficient.

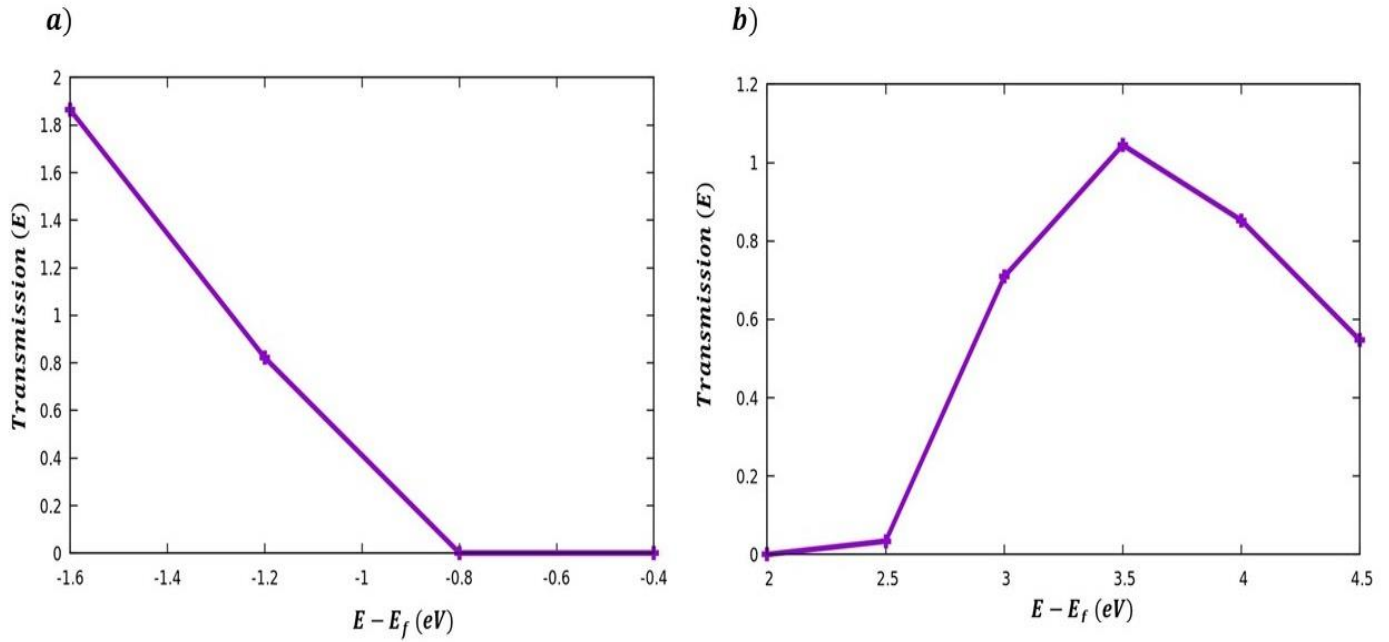

*Figure SI.6. Transmission analysis for energies far from Fermi level of $8+8\ WSe_2//NbS_2\ LH$: a) below the Fermi level, and b) above the Fermi level.*